# Multi-source data processing and fusion method for power distribution internet of things based on edge intelligence

**Quande Yuan[1], Yuzhen Pi[1,3,*], Lei Kou[2], Fangfang Zhang[2], Yang Li[4], Zhenming Zhang[4]**

[1]Changchun Institute of Technology, Changchun, China

[2]Qilu University of Technology (Shandong Academy of Sciences), Qingdao, China

[3]National Local Joint Engineering Research Center for Smart Distribution Grid Measurement and Control with Safety Operation Technology, Changchun Institute of Technology, Changchun, China

[4]Northeast Electric Power University, Jilin, China

**\* Correspondence:**
Yuzhen Pi
piyz@airlab.ac.cn

## Abstract

With the rapid advancement of the Energy Internet strategy, the number of sensors within the Power Distribution Internet of Things (PD-IoT) has increased dramatically. In this paper, an edge intelligence-based PD-IoT multi-source data processing and fusion method is proposed to solve the problems of confusing storage and insufficient fusion computing performance of multi-source heterogeneous distribution data. First, a PD-IoT multi-source data processing and fusion architecture based on edge smart terminals is designed. Second, to realize the uniform conversion of various sensor data sources in the distribution network in terms of magnitude and order of magnitude. By introducing the Box-Cox transform to improve the data offset problem in the Zscore normalization process, a multi-source heterogeneous data processing method for distribution networks based on the Box-Cox transform Zscore is proposed. Then, the conflicting phenomena of DS inference methods in data source fusion are optimally handled based on the PCA algorithm. A multi-source data fusion model based on DS inference with conflict optimization is constructed to ensure the effective fusion of distribution data sources from different domains. Finally, the effectiveness of the proposed method is verified by an experimental analysis of an IEEE39 node system in a regional distribution network in China.

## 1    Introduction

In the power system, along with the intelligent advancement of the Power Distribution Internet of Things (PD-IoT). Millions of electrical quantity sensors, power distribution devices and condition sensors will be connected to the IoT network (Motepe et al., 2019; Yin et al., 2020). It generates a huge amount of heterogeneous distribution data, and presents a wide variety, multiple sources, and uncertainty. The traditional grid center is used to realize the cloud-edge computing mode. Due to the rapid increase of massive data and the influence of data complexity, the upstream terminal monitoring data and power grid operation data as well as the downstream cloud computing processing information bring great pressure to the communication transmission layer, and seriously restrict the promotion of PD-IoT (Zhang et al., 2018). In depth exploration of the application potential of edge computing has become a major research focus, and more and more distribution network data information terminal processing, marginal computing and local solutions have become an important way (Khalifa et al., 2018; Luo et al., 2021). However, the marginalization integration of complex multi-source heterogeneous data brings new challenges to efficient edge computing in the PD-IoT mode. Therefore, it is urgent to realize the processing and fusion of power distribution internet of things multi-source data in the marginalization mode, which is an important basis for improving and ensuring the intelligent development of power distribution internet of things based on edge computing (Feng et al., 2020; Liu et al., 2021).

For a long time, the PD-IoT huge amounts of data analysis of mining have been taken into account in some previous works by using previous methods, Dashtdar et al. (Dashtdar et al., 2021) for distribution network, such as data source and data analysis, and designed a contains four levels of distribution network operation data analysis system architecture; Sun Liu et al. (Liu et al., 2020; Li et al., 2021a) proposed an application method of edge computing technology in the internet of things of distribution network, and deeply analyzed the shortcomings of the practical application of edge computing technology in the distribution industry. Han et al. (Han et al., 2021) proposed an application method of edge computing architecture in smart power grid model, and combined with intelligent terminal data and operational measurement data, elaborated the significance of edge computing in data security and efficiency analysis. Wang et al. (Kou et al., 2020; Wang et al., 2021) proposed a strategy of assisting distribution network fault information identification and location with the analysis results of multi-source heterogeneous data, which provided a research reference for effectively mining the application value of intelligent data of distribution network. Merad-Boudia et al. (Merad-Boudia et al., 2020) proposed to enhance smart grid by integrating advanced metering infrastructure and fog computing, extending distributed control, communication and computing capabilities, and improving the reliability, flexibility and scalability of smart grid. Sahu et al. (Sahu et al., 2021) established an adaptive multi-objective group cross optimization algorithm to achieve the classification fusion of multi-source data and accurate fusion of heterogeneous data.

To sum up, power distribution internet of things has become an inevitable trend in the development of power system distribution side. The application of edge computing technology provides distributed services (Li et al., 2021b) and computing functions, but the research on effective cleaning and connection of massive multi-source heterogeneous grid data is insufficient, so the huge potential of edge computing performance cannot be brought into full play. In this paper, a multi-source data processing and fusion method of power distribution internet of things based on edge intelligence is proposed. It can effectively realize the fusion of distribution network operation data, terminal monitoring data, environmental information data and other basic data sources, and lay a good foundation for enhancing the multi-source parallel mining and fusion computing analysis of new power system big data, which has important research value and significance.

The main contributions of this paper are as follows:

(i) Design a power distribution internet of things data processing and convergence architecture that fully considers edge computing.

(ii) A unified processing method based on Box-Cox transformation Zscore (BC-Zscore) is proposed for the dimensionality and order of magnitude transformation of data sources in power distribution networks.

(iii) By constructing a multi-source data fusion model based on Principal Components Analysis-Dempster Shafer (PCA-DS), the multi-source heterogeneous data were grouped and aggregated based on multidimensional feature factors.

This paper is organized as follows: Section 2 provides a brief discussion of the data processing and fusion architecture. Among them, the multi-source data normalization processing method is described in detail in Section 3. Section 4 provides the process for building a multi-source data fusion model for the power distribution network. Experimental comparison results and performance analysis of the proposed method are given in Section 5. Finally, conclusions and future recommendations are contained in Section 6.

## 2 Power Distribution Internet of Things Multi-source Data Processing and Fusion Architecture

The power distribution side of the power system is connected with the power transmission system through the distribution substation, which is the last link to transfer the power resources from the transmission system to the users (Shen et al., 2019; Qu et al., 2021). In the process, the medium voltage grade power is transmitted to the distribution transformer located near the user's office by the primary distribution line. The power is then reduced again by distribution transformers to the use voltage of lighting, industrial equipment and household appliances; Finally, the secondary distribution line supplies power to the associated users, and the electricity consumption of customers is recorded through the electricity meter. The construction of PD-IoT realizes the monitoring, protection and control of the entire power distribution system under stable and abnormal operation by introducing cutting-edge technologies in the fields of modern electronics, communications, networks and computers (Jiang et al., 2020).

Combined with edge computing definition and technical features, it can effectively solve the core link in the construction process of power distribution IoT. It acts between the end power operation equipment and intelligent monitoring equipment and the cloud master station to realize the basis of data aggregation, data computing, data storage and higher-level data application. It gives full play to the edge structure advantage of local computing to achieve the goal of power distribution business function of PD-IoT terminal expansion, topology flexibility and real-time counting and control (Chen et al., 2021; Zhong et al., 2021). The realization of this goal or the degree of realization depends on the data aggregation operation under the big data of electricity distribution. The aggregation of electricity distribution data is not only the aggregation and integration of data, but also the processing transformation of multiple sources and levels of heterogeneous information data in the distribution management system, as well as the deep integration that fully considers the association of characteristics and attributes among different data sources. Efficient multi-source data processing and fusion can make power distribution data serve edge computing better and improve the application ability of edge computing in power distribution internet of things. The multi-source data processing and fusion architecture of power distribution internet of things considering edge computing is shown in **Figure 1**.

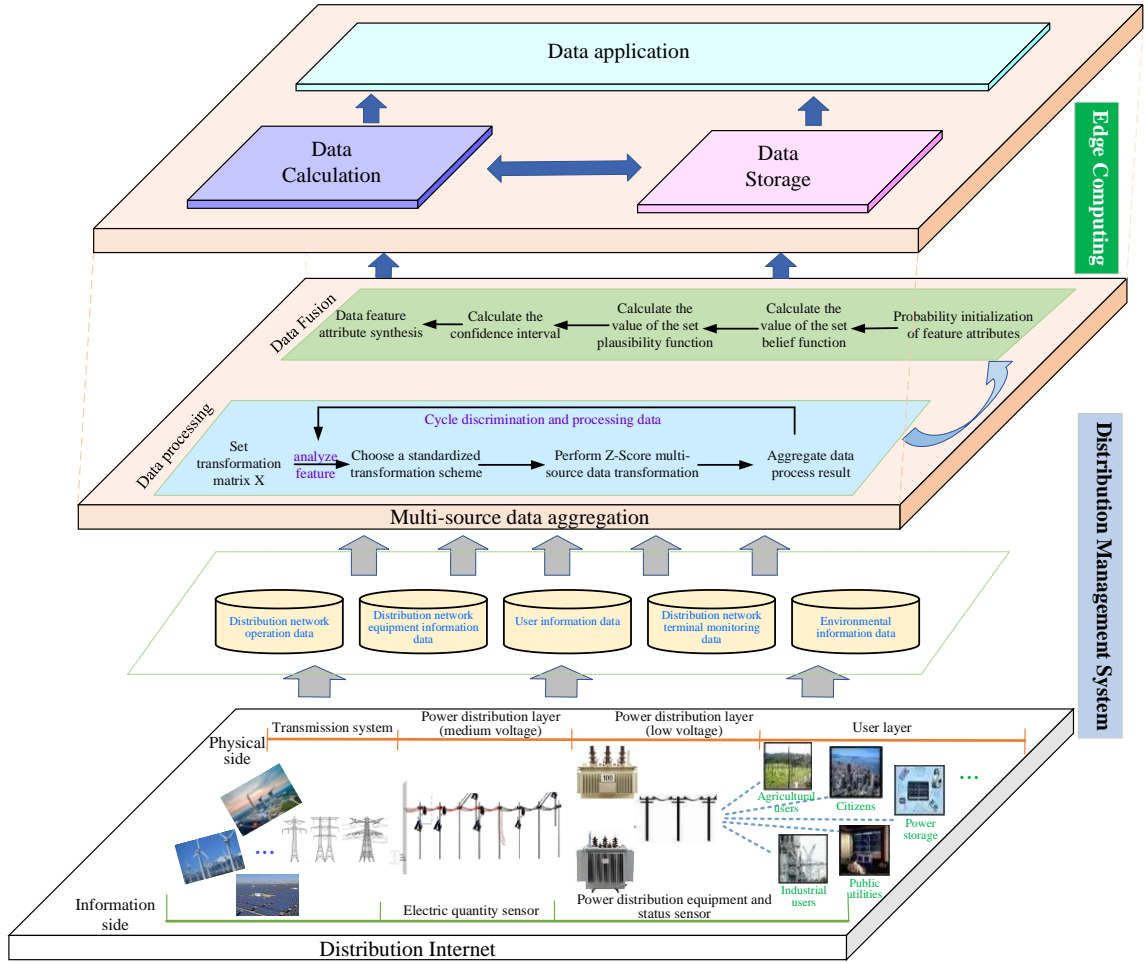

**Figure 1** Multi-source data processing and fusion architecture.

## 3    Power Distribution Internet of Things Multi-source Data Standardized Processing

In the process of distribution network operation, the format, dimension, data type and order of magnitude of feature attributes of various data sources are different. In order to realize the data fusion calculation and information mining of distribution internet of things marginalization. It is necessary to eliminate the restrictions caused by various inconsistent factors and realize the standardized processing of multi-source data. According to the temporal characteristics of distribution big data, Box-Cox transformation is introduced on the basis of the original Zscore multi-source heterogeneous data standardized processing. A method of power distribution network multi-source heterogeneous data standardized processing based on Box-Cox transformation Zscore is proposed in this paper.

Since Zscore data standardization is to assume that the multi-source heterogeneous data factor to obey the law of normal distribution, otherwise the time-series data of skewness and kurtosis makes data processing in the process of the influence of a certain factor scores on small or large (Khond et al., 2020; Kou et al., 2021). The Box-Cox transformation, as a generalized power transformation method, can effectively handle the case where the continuous response variables do not satisfy the normal distribution. It can eliminate the problem of fluctuating offset of multi-source timing data generated during the operation or collection of distribution data, and ensure the accuracy and stability of standardized processing of multi-source heterogeneous data. The specific processing process is shown in **Figure 2**.

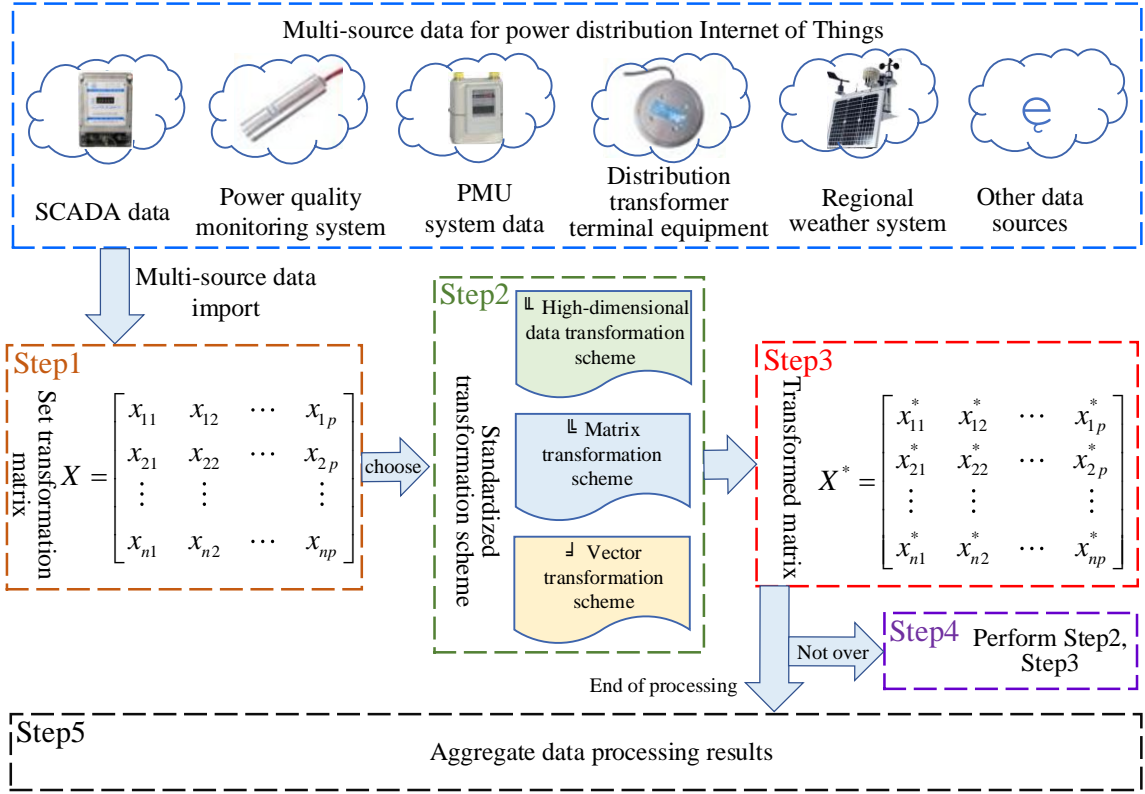

**Figure 2** Multi-source data standardization process.

The detailed procedure is as follows.

Step 1: Decompose and process the collected multi-source data of power distribution internet of things according to the time sequence characteristics, and write $X$ as the standardized transformation input, where $X$ can exist in the form of multidimensional data, matrix and vector. When $X = \left( X_1, X_2, \ldots, X_p \right)$ in matrix form:

$$X = \begin{bmatrix} x_{11} & x_{12} & \cdots & x_{1p} \\ x_{21} & x_{22} & \cdots & x_{2p} \\ \vdots & \vdots & & \vdots \\ x_{n1} & x_{n2} & \cdots & x_{np} \end{bmatrix} \quad （1）$$

Step 2: Due to the characteristics of multi-source heterogeneity of power distribution big data, BC-Zscore standardized transformation schemes of different formats are set respectively to ensure that data transformation processing can be achieved in different source data formats:

If $X$ exists in the form of multi-dimensional array, the mean and standard deviation are solved along multiple dimensions of $X$, and then the data of $X$ is normalized to return the transformed high-dimensional array $BC\_Z = (X - mean(X)) / std(X)$;

If $X$ exists in vector form, the result vector after transformation is returned BC_Z;

If $X$ exists in the form of matrix, the mean and standard deviation of the column vectors of $X$ are used to conduct data normalization processing for the corresponding columns one by one, and the resulting matrix BC_Z is returned.

Step 3: $X = \left( X_1, X_2, \ldots, X_p \right)$ standardized processing of BC-ZScore data:

$$X^* = \begin{bmatrix} x_{11}^* & x_{12}^* & \cdots & x_{1p}^* \\ x_{21}^* & x_{22}^* & \cdots & x_{2p}^* \\ \vdots & \vdots & & \vdots \\ x_{n1}^* & x_{n2}^* & \cdots & x_{np}^* \end{bmatrix} \quad （2）$$

where $x_{ij}^* = \dfrac{x_{ij} - \overline{x}_j}{\sqrt{s_{ij}}}, i = 1, 2, \cdots, n, j = 1, 2, \cdots, p$. Among them $\overline{x}_j = \dfrac{1}{n} \sum_{i=1}^{n} x_{ij}$ solve the average value of variables $X_j$; $\sqrt{s_{ij}} = \sqrt{\dfrac{1}{n-1} \sum_{i=1}^{n} (x_{ij} - \overline{x}_j)^2}$ find the standard deviation of the variable $X_j$. After BC-Zscore data transformation processing $X = (X_1, X_2, \cdots, X_p)$ each of the columns $\sqrt{s_{ij}} = 1, \overline{x}_j = 0$, $j = 1, 2, \cdots, p$.

Step 4: Perform data iterative processing according to the selected multi-source data processing scheme, perform step 2 and step 3 in a cycle, and gather data transformation processing results.

Step 5: After all the data sources of task input undergo the unified dimension and magnitude transformation, the task output is saved for multi-source data fusion, edge data calculation or data storage, and the standardization transformation of multi-source data is completed.

## 4 Building a Multi-source Data Fusion Model Based on Conflict-optimized DS Inference

Power distribution internet of things based on edge computing mode mainly includes three stages: information fusion, state evaluation and associated decision. Information fusion is realized through the processing and fusion of big data of power distribution, which only involves a few brief data calculations, and the calculation of key data is in the stage after information fusion (Krishnamurthi et al., 2020; Lau et al., 2017). Therefore, the performance and significance of data fusion are very important, which is directly related to the power distribution network state evaluation and associated decision calculation results. Based on the standardized processing of multi-source data, this paper constructs the multi-source data fusion model of distribution network based on DS reasoning of conflict optimization. Data fusion combines heterogeneous data from multiple data sources or from related databases to achieve higher accuracy than edge computing using a single data source.

The Dempster-Shafer (DS) inference method, as a classical data fusion method for dealing with uncertainty, achieves a further improvement on Bayesian conditional probability in probability theory. It avoids the calculation of prior probabilities and can represent "uncertainty" well, which is widely used in various fields of data fusion (Jing et al., 2021). However, when DS inference method deals with conflict subsets, the normalization process of combination rules will violate the common sense of fusion of different data sources. PCA algorithm (Li et al., 2022) is applied to further optimize DS inference method when dealing with conflict data source fusion. It can achieve the goal of finding m (m< n) new components, make them reflect the main characteristics of conflict information, and realize the extraction and utilization of the main components of conflict information, instead of assigning all the components to unknown terms without considering fusion. The available components depend on the defined component reliability function:

$$\tilde{k} = \sum_{k=1}^{n} e_i BV_{i,k} \qquad (3)$$

where $\tilde{k}$ reflecting the degree of conflict between the components; $e_i$ represents different conflict components in conflict information; The $BV$ is the value of the corresponding dimension in the conflicting data, $i$ represents the dimension of the conflicting data, $i = 1, 2, \cdots, m$.

In view of the diverse, multi-source and uncertain characteristics of electricity distribution big data, the feature-level fusion of multi-source heterogeneous data under the PD-IoT is realized by abstracting the data sources or monitoring terminal data information into feature attribute subsets. The specific implementation steps are as follows.

Step 1: Initialize the basic probability of the subset of multi-source feature attributes, mark $U$ is the multi-source data fusion model framework of power distribution internet of things, then function $m : 2^U \to [0,1]$ satisfies two conditions:

$$\begin{cases} m(a) = 0 \\ \sum_{A \subset U} m(a) = 1 \end{cases} \qquad (4)$$

where $m(a) = 0$ the initial value of multi-source data fusion set A, and the size of $m(a)$ represents the degree of trust in it.

Step 2: Define belief function to calculate the trust function value of different data fusion sets.

$$Bel : 2^u \to [0,1]$$
$$Bel(A) = \sum_{a \subset A} m(a), (\forall A \subset U) \qquad (5)$$

where $Bel(A)$ represents the sum of the distribution probability values of all subsets in the multi-source data fusion set $A$, and each distribution probability value represents the trust degree value of the feature attributes of the sub-set, indicating that the multi-source feature attributes of the power distribution network included in it can realize the most basic data fusion.

Step 3: Define the plausibility function to calculate the trust degree value of the fusion uncertain feature attribute set. The available component of uncertain feature attribute depends on the reliability value $\tilde{k}$ of the component to be solved. The calculation Function is as follows:

where $pl(A)$ represents the measure of the uncertain characteristic attributes of multi-source data fusion set $A$ that seems likely to be fused; $\tilde{k}_{A \cap B}$ denotes the conflicting component confidence value when $A$ fuses uncertainty feature attributes, and is obtained by the calculation of equation (3).

Step 4: Calculate the trust space for data fusion. According to the relationship between trust function and likelihood function: $pl(A) \geq Bel(A), A \subset U$, the uncertainty of $A$ can be expressed as:

$$\mu(A) = pl(A) - Bel(A)$$

（7）

where $(pl(A) - Bel(A))$ is the trust space, represents the uncertain characteristic attributes that are allowed to change according to the actual application of distribution network calculation in the process of multi-source data fusion.

Step 5: Multi-source heterogeneous data feature attribute synthesis. For $\forall A \subset U$, dempster's synthesis rule for the limited Mass functions $m_1, m_1, \cdots, m_n$ on the multi-source data fusion model framework $U$ of power distribution internet of things is:

$$(m_1 \oplus m_2 \oplus \cdots \oplus m_n)(A) = \frac{1}{K} \sum_{A_1 \cap A_2 \cap \cdots \cap A_n = A} m_1(A_1) \bullet m_2(A_2) \cdots m_n(A_n)$$

（8）

where $K$ is expressed as $\sum_{A_1 \cap A_2 \cap \cdots \cap A_n \neq \varnothing} m_1(A_1) \bullet m_2(A_2) \cdots m_n(A_n)$, according to the rule of composition, the feature attribute index of data from different sources is used to realize the feature level data fusion.

## 5    Experimental Results and Analysis

To verify the effectiveness and reliability of the proposed edge computing-based multi-source data processing and fusion technique for PD-IoT in the paper. Taking a regional distribution network in China as an example, the topology is assumed to be the actual topology modified IEEE39 node system. The algorithm is implemented by Matlab2020a. The power distribution network operation data, terminal monitoring data and environmental information data in the same time cycle are selected as experimental multi-source data. The algorithm is divided into two experiments to test the multi-source data standardization transformation processing method and the feasibility of multi-source data feature level fusion based on data processing.

**Experiment 1**: Multi-source data standardized transformation processing. Operation measurement of distribution network mainly uses data of Supervisory Control and Data Acquisition (SCADA); Terminal monitoring data mainly uses System of Fault Distribution (SMD) data. The relevant environmental information data is publicized by meteorological network, mainly including time point, longitude, latitude, temperature, wind speed, air pressure and other characteristic attribute data that have significant influence on distribution network. Each data source belongs to a different data acquisition system, and the data format is complex and diverse with great structural differences, which greatly limits the marginal mixed computing of distribution network. The comparison of results before and after multi-source data standardized transformation is shown in **Figure 3A** and **Figure 3B**. In order to avoid the contingency of the processing method, 10 observation serial numbers were randomly selected for 6-11 lines as an example to show the results. After being processed by the method, the processing results of three data source characteristic attributes with different formats, dimensions, data types and orders of magnitude are shown in **Table 1**.

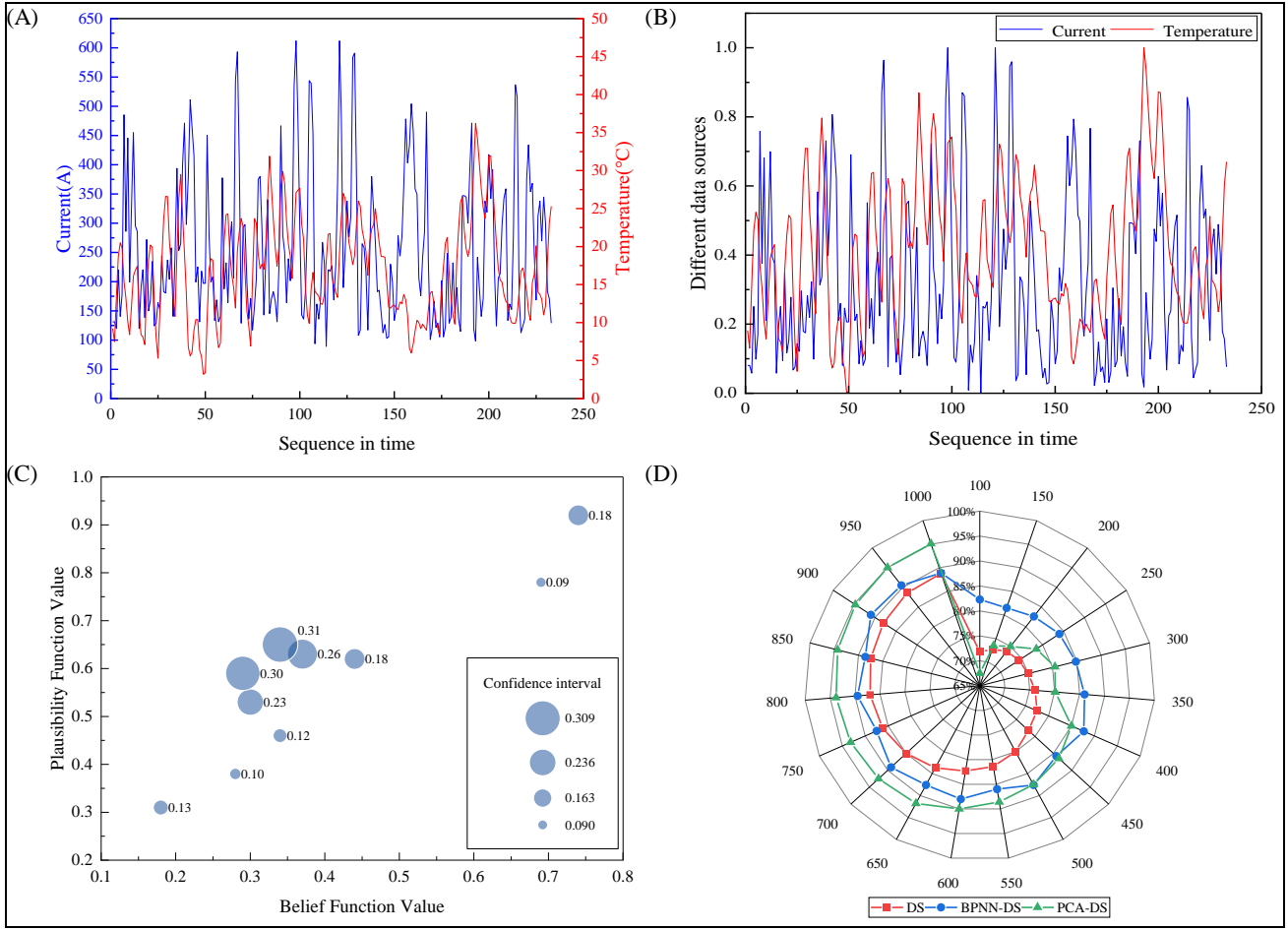

**Figure 3** (A) Results before multi-source data processing. (B) Results after multi-source data processing. (C) Change of confidence interval during data fusion. (D) Comparison of the accuracy of the results of the three algorithms in the later diagnosis analysis.

**Table 1** BC-Zscore multi-source data processing experimental results values

| Observation number | Terminal monitoring data | | Environmental information data | | Distribution network operation data | |
| --- | --- | --- | --- | --- | --- | --- |
| | Electric energy | Power factor | Temperature | Wind speed | Line voltage | Line current |
| 1 | 1.00 | 0.25 | 0.42 | 0.84 | 0.04 | 0.33 |
| 2 | 0.72 | 0.49 | 0.38 | 0.28 | 0.38 | 0.52 |
| 3 | 0.75 | 0.37 | 0.53 | 0.30 | 0.06 | 0.60 |
| 4 | 0.01 | 0.22 | 0.92 | 0.74 | 0.24 | 0.81 |
| 5 | 0.56 | 0.80 | 0.78 | 0.99 | 0.17 | 0.93 |
| 6 | 0.32 | 0.07 | 0.66 | 0.58 | 0.54 | 0.92 |
| 7 | 0.58 | 0.18 | 0.63 | 0.37 | 0.94 | 0.26 |
| 8 | 0.16 | 0.04 | 0.34 | 0.28 | 0.98 | 0.41 |
| 9 | 0.44 | 0.72 | 0.59 | 0.29 | 0.43 | 0.27 |
| 10 | 0.14 | 0.72 | 0.65 | 0.34 | 0.65 | 0.03 |

By standardizing the continuous original data within a week based on BC-Zscore for multi-source data, this method effectively achieves the unified transformation of the format, dimension, data type and order of magnitude of the feature attributes of each data source. **Table 1** shows the experimental results. The absolute value of each element is between 0 and 1 after the characteristic attribute transformation of each data source, which obviously eliminates the limitation caused by various inconsistent factors, lays a foundation for the subsequent marginal data fusion calculation and information mining of distribution Internet of Things, and ensures the stable calculation control of distribution network and the in-depth mining of important information.

**Experiment 2**: Multi-source data fusion. All kinds of advanced application computing of power distribution internet of things are based on multi-source data fusion computing. Based on multi-source data processing results, feature level fusion of data is carried out centering on the actual application scenario demand of natural disaster fault information diagnosis and mining of distribution network. Taking operation data sources, terminal monitoring data sources and environmental information data sources as objects, the composition and key characteristic attributes of multi-source information are analyzed in combination with the requirements of application scenarios.

Classification and fusion of multi-source data according to the rules for synthesizing feature attributes of multi-source heterogeneous data. The changes of Confidence interval under different Belief Function Value and Plausibility Function Value for 10 randomly selected observation serial numbers in **Table 1** during data fusion are shown in **Figure 3C**. It finds the best data fusion point and fuses as many uncertain feature attributes as possible with guaranteed stability.

In order to verify the high accuracy of the multi-source data fusion results of PCA-DS reasoning proposed in this paper in the diagnosis and mining of natural disaster fault information of power distribution network in the later stage. For the experimental result values of 1000 records processed by BC-ZScore multi-source data within a week, DS reasoning method, BP neural network combined with DS reasoning algorithm (BPNN-DS) and the algorithm in this paper (PCA-DS) were used to compare the fusion effect. The accuracy changes of the fusion results of the three experiments in later diagnostic analysis are shown in **Figure 3D**.

Experimental results show that the multi-source data fusion model based on PCA-DS reasoning can effectively achieve the grouping and aggregation of multi-source heterogeneous data from different perspectives such as data source and characteristics. According to the contraction of trust interval in the fusion process, the fitting accuracy of distribution characteristics of multi-source data is improved. At the same time, through the comparison results of distribution network natural disaster fault information diagnosis and mining considering multi-source data fusion, it can be seen that the stability and accuracy of DS reasoning in multi-source data fusion are greatly improved by using the defined component reliability function to constrain the fusion of uncertain feature attributes in DS reasoning process. Moreover, the multi-source information synthesis under the effective fusion method is obviously superior to the traditional method which only considers a single or a few factors.

## 6    Conclusions

In order to solve the problems of storage confusion and insufficient fusion computing performance caused by massive heterogeneous data in the process of intelligent construction of distribution network, this paper proposes a multi-source data processing and fusion technology for PD-IoT based on edge computing. The proposal of this technology has accelerated the application of edge computing technology in the PD-IoT to a certain extent. The simulation results of an actual regional distribution network show that:

1) The design of a power distribution data processing and fusion architecture that fully considers the edge computing mode provides a guarantee for the subsequent advanced computing and application decision analysis of the distribution network.

2) The proposed generalized power transformation Zscore multi-source data transformation processing method effectively realizes the unification of dimension and magnitude under various distribution network data acquisition systems.

3) Based on the data fusion model of conflict optimization DS reasoning, high-precision feature-level power consumption data fusion is realized in advance according to the requirements of advanced application scenarios of the PD-IoT. In addition, the model lays the foundation for intelligent computing and advanced applications at the edge of the distribution grid.

In the future work, we will extend this method to the specific business applications of the power grid to meet the needs of advanced applications such as operation status assessment, fault information diagnosis and mining, and emergency data reliability identification of the PD-IoT. Furthermore, it would be an interesting topic to explore more multi-source data fusion in the context of renewable energy.

## 7    Data Availability Statement

The raw data supporting the conclusions of this article will be made available by the authors, without undue reservation.

## 8    Author Contributions

QY: designed this study. YP: contributed to the power distribution internet of things multi-source data processing and fusion architecture. LK: contributed to the power distribution internet of things multi-source data standardized processing. FZ: collected and cleansed the data. YL: built a multi-source data fusion model based on conflict-optimized DS inference. ZZ: completed data cleaning and carried out detailed experimental analysis. All authors contributed to the writing of the article and all agreed to the submitted version of the article.

## 9    Funding

This paper was partly supported by the Science and Technology Development Plan of Jilin Province (grant numbers: 20210201049GX), and the Science and Technology Projects of Education Department of Jilin Province (grant numbers: JJKH20191262KJ, JJKH20191258KJ).

## 10    Conflict of Interest

The authors declare that the research was conducted in the absence of any commercial or financial relationships that could be construed as a potential conflict of interest.